\documentclass[twocolumn]{aastex62}

\newcommand{\pder}[2]{\frac{\partial#1}{\partial#2}}

\newcommand{\der}[2]{\frac{d#1}{d#2}}
\newcommand{\dder}[2]{\frac{d^2#1}{d#2^2}}

\newcommand{\Fs}{\mathcal{F}_*}

\newcommand{\ve}{\varepsilon}
\newcommand{\ves}{\varepsilon_s}
\newcommand{\vecm}{\varepsilon_c}
\newcommand{\vel}{\varepsilon_l}
\newcommand{\kappac}{\kappa_c}
\newcommand{\kappal}{\kappa_l}
\newcommand{\kc}{k_c}
\newcommand{\kl}{k_l}
\newcommand{\gamu}{\gamma_{\mu}}
\newcommand{\gamus}{\gamma_{\mu \varepsilon}}
\newcommand{\calf}{\mathcal{F}}
\newcommand{\calh}{\mathcal{H}}
\newcommand{\tB}{\mathcal{B}}

\newcommand{\teta}{\tilde{\eta}}
\newcommand{\Tint}{T_{\mathrm{int}}}
\newcommand{\Tirr}{T_{\mathrm{irr}}}
\newcommand{\tl}{\tilde{\lambda}}

\submitjournal{ApJ}

\shorttitle{Picket Fence Model of Irradiated Atmospheres with
Coherent Scattering}
\shortauthors{Mohandas et al.}

\begin{document}

\title{Analytical Models of Exoplanetary Atmospheres. V. Non-gray Thermal
Structure with Coherent Scattering}

\correspondingauthor{Gopakumar Mohandas}
\email{gopakumar@nbi.ku.dk}

\author[0000-0002-1212-3492]{Gopakumar Mohandas}
\affiliation{Niels Bohr International Academy,
Niels Bohr Institute,
Blegdamsvej 17, 2100 Copenhagen, Denmark}
\affiliation{Kavli Institute for Theoretical Physics,
University of California, Santa Barbara 93106, USA
}

\author{Martin E. Pessah}
\affiliation{Niels Bohr International Academy,
Niels Bohr Institute,
Blegdamsvej 17, 2100 Copenhagen, Denmark}

\author{Kevin Heng}
\affiliation{University of Bern,
Center for Space and Habitability,
Gesellschaftsstrasse 6, CH-3012, Bern, Switzerland}



\begin{abstract}
We apply the picket fence treatment to model the effects brought about by
spectral lines on the thermal structure of irradiated atmospheres.
The lines may be due to purely absorption processes, purely coherent
scattering processes or some combination of absorption and scattering.
If the lines arise as a pure absorption process, the surface layers of the atmosphere are cooler whereas this surface cooling is completely absent if the lines are due to pure coherent isotropic scattering. The lines also
lead to a warming of the deeper atmosphere. The warming of the deeper layers
is, however, independent of the nature of line formation.
Accounting for coherent isotropic scattering in the shortwave and
longwave continuum results in anti-greenhouse cooling and greenhouse warming
on an atmosphere-wide scale.
The effects of coherent isotropic scattering in the line and continuum operate
in tandem to determine the resulting thermal structure of the irradiated atmosphere.
\end{abstract}



\section{Introduction}

Analytical radiative transfer models have proved to be of considerable utility
in the study of stellar and planetary atmospheres since their development a
century ago \citep{mihalas70}.
Despite the availability of sophisticated high-speed numerical techniques
in the present day \citep{hubeny2017},
simplified analytical treatments continue to remain valuable primarily as
a means to derive physical insight and understanding of atmospheric
conditions.

The usefulness of analytical models to construct model
atmospheres are afforded by the simplistic nature of the underlying
assumptions, namely that of gray opacities,
plane-parallel steady-state structure, hydrostatic and radiative equilibrium
\citep{mihalas70}. These assumptions are however also the source of their
limitations. In particular, the assumption of frequency independent (gray)
mean opacities are far from a realistic representation of true atmospheric
opacities \citep{hengbook}.
Nevertheless, such simplified 1D analytical models have been useful in
providing an exact solution that predicts basic atmospheric trends
and one that may serve as a good initial approximation for numerical schemes
\citep{hubenymihalas}.

Analytical or semi-analytical models with small departures from grayness have
been derived over the years \citep{hubenymihalas}. In the
context of irradiated atmospheres, such as that of close-in extra solar
planets, the most elementary extension constitutes what is referred to as the
\emph{two-step gray} or \emph{semi-gray} or \emph{dual-band} transfer models
\citep{hansen2008,guillot2010,hengetal2012,hengetalrt2014}. These models are
predicated on the approximate division of radiant energy into two distinct and
nearly non-overlapping bands; the shortwave associated with external
stellar irradiation and the longwave associated with internal planetary
thermal emission. The transfer equations or their moments, with different mean
opacities for the shortwave and longwave components, are then solved
separately and linked together by the principle of energy conservation.

Recently, \citet{pg2014} derived an analytical model by applying the
\emph{picket fence} method of \citet{chandra35} to irradiated atmospheres.
The picket fence treatment was originally developed to model line blanketing
effects in non-irradiated stellar atmospheres and has since been
refined by a number of authors \citep{munch46,ml1971,as1969}. Spectral line
blanketing leads to two major effects that introduces departures from a gray
atmosphere: surface cooling and backwarming \citep{ml1971,athaybook}.
The former refers to the reduction of temperature in the upper layers of the
atmosphere due
to the added emissivity of the lines whereas the latter effect describes
the temperature enhancement in the deeper atmosphere as a result of the flux
redistribution within the continuum due to the lines.
Both these effects were present in the model derived by \citet{pg2014}.
However, their analysis did not account for the influence of scattering in
both the shortwave and the longwave. The surface cooling effect
has been seen to be dependent on the nature of line formation in
non-irradiated atmospheres \citep{chandra35}. The degree of cooling is
lower if scattering contributes to the line in some measure and is
completely absent when the lines are entirely due to pure scattering process.
One would expect this feature to be present in irradiated atmospheres
as well. Furthermore, continuum scattering is known to induce an atmosphere
wide shift in temperatures \citep{hengetalrt2014,hengbook}.
This shift is towards hotter temperatures if the isotropic scattering
contributes to the longwave continuum and is towards lower temperatures if
isotropic scattering contributes to the shortwave continuum.
Therefore, one must account for scattering in the line and continuum in
order to derive a closer approximation to actual atmospheric thermal
structures.

In this paper, we generalize the picket fence treatment to irradiated
atmospheres to include coherent scattering effects.
We derive solutions that accommodate for coherent isotropic scattering
in the lines as well as the continuum, in the longwave and shortwave
frequency bands.
Our model therefore provides a fuller picture of the
possible atmospheric thermal structure while preserving the advantages and
utility rendered by tractable non-gray analytical models.

The paper is organized as follows. In Section~\ref{sec:equations}, we
present the picket fence model equations. In Section \ref{sec:solutions},
we connect the longwave picket fence model with the shortwave equations and
solve for the resulting temperature profile. We present a discussion of
the results in Section \ref{sec:discussion} and conclude with a summary in
Section \ref{sec:summary}.

\section{The Picket Fence Model}
\label{sec:equations}

We consider a plane-parallel atmosphere and model the transport of radiation
by solving the moments of the steady-state radiative transfer equation.
We apply the dual-band approximation and derive separate moment equations in
the longwave and shortwave frequency band.

We begin with the longwave band where we make use of the picket fence
method \citep{chandra35}. The radiative transfer equation in a plane-parallel
atmosphere has the basic form
\begin{equation}
  \mu\pder{I_{\nu}}{m} = k(\nu)(I_{\nu} - S_{\nu}),
\end{equation}
where $\mu$ is the cosine of the zenith angle, $m$ is the column mass,
$k$ is the frequency dependent extinction opacity,
$I_{\nu}$ is the specific intensity and $S_{\nu}$ is the source function.
The intensity and source function are in general functions of $\mu$, $m$
and $\nu$.

We define the moments of the intensity; the mean intensity, the Eddington flux
and the $K$-integral respectively as follows \citep{mihalas70}
\begin{eqnarray}
\label{eq:J}
  J_{\nu} &\equiv& \frac{1}{2}\int_{-1}^{1} I_{\nu}(\tau, \mu) d\mu,  \\
\label{eq:H}
  H_{\nu} &\equiv& \frac{1}{2}\int_{-1}^{1} I_{\nu}(\tau, \mu) \mu d\mu,  \\
\label{eq:K}
  K_{\nu} &\equiv& \frac{1}{2}\int_{-1}^{1} I_{\nu}(\tau, \mu) \mu^2 d\mu.
\end{eqnarray}

In the parts of the frequency interval containing lines, the equation of
transfer is \citep{athaybook,mihalas70}
\begin{equation}
  \mu\pder{I_{\nu}}{m} = \left(k_c + k_l\right)I_{\nu} -
  \left(k_c S_{c\nu} + k_l S_{l\nu} \right),
\end{equation}
where the subscripts $c$ and $l$ refer to the continuum
and line respectively. Considering coherent isotropic scattering,
the continuum and line source functions have the form
\begin{eqnarray}
  S_{c\nu} &=& \vecm B_{\nu} + (1 - \vecm) J_{\nu}, \quad \vecm =
  \frac{\kappac}{\kc}, \\
  S_{l\nu} &=& \vel B_{\nu} + (1 - \vel)J_{\nu}, \quad
  \vel = \frac{\kappal}{\kl},
\end{eqnarray}
where $\kappa$ denotes the absorption opacity that together with the
corresponding scattering opacity assume constant but separate values
in the line and continuum. The parameter $\varepsilon$
is a measure of the fraction of photons lost to pure absorption \citep{mihalas70}.
It is in fact the complementary parameter to the single scattering
albedo but we shall refer to it here as the scattering parameter regardless.

Integrating over the frequency interval containing only the lines leads to the
transfer equation
\begin{eqnarray}
\label{eq:line-intensity}
  \mu\pder{I_1}{\tau} &=&
  \left(\frac{1}{\vecm} + \xi \right)I_1
  - \left[ \frac{1}{\vecm} - 1 + (1 - \vel)\xi \right] J_1
  \nonumber \\ &\phantom{=}&
  - (1 + \vel\xi)\beta\tB,
\end{eqnarray}
whereas integrating over the remainder of the frequency interval representing
the continuum yields
\begin{equation}
\label{eq:cont-intensity}
  \vecm\mu\pder{I_2}{\tau} = I_2 - (1 - \vecm)J_2 - (1 - \beta)\vecm\tB.
\end{equation}
Here the subscripts $1$ and $2$ represent the integrated variables in the
line and continuum respectively.
We have also defined the line to continuum opacity ratio
\begin{equation}
  \xi = \frac{k_l}{\kappa_c},
\end{equation}
the integrated blackbody function
\begin{equation}
	\tB = \int B_{\nu} d\nu,
\end{equation}
and the frequency independent optical depth
\begin{equation}
	d\tau = \kappa_c dm.
\end{equation}
Finally, we introduce the parameter $\beta$ which gives the relative probability
of finding a line in the frequency interval and may in general be a function of frequency.
However, we take $\beta$ to be a constant in Equations~(\ref{eq:line-intensity})
and (\ref{eq:cont-intensity}) assuming that the lines have uniform width and are
uniformly spread across the spectral range.

The first and second moments of the radiative transfer equation in the line and
continuum are respectively
\begin{eqnarray}
\label{eq:firstmom-1}
  \der{H_1}{\tau} &=& \lambda(J_1 - \beta\tB), \\
\label{eq:firstmom-2}
  \der{H_2}{\tau} &=& J_2 - (1 - \beta)\tB,
\end{eqnarray}
and
\begin{eqnarray}
\label{eq:secmom-1}
  \der{K_1}{\tau} &=& \frac{\eta}{\vecm} H_1, \\
\label{eq:secmom-2}
  \der{K_2}{\tau} &=& \frac{1}{\vecm} H_2,
\end{eqnarray}
where we have defined
\begin{equation}
  \lambda = 1 + \vel\xi, \quad \eta = 1 + \vecm\xi.
\end{equation}
In the limit $\vecm = 1$, we recover the moment equations used
by \citet{chandra35}.

\section{Irradiated Atmospheres}
\label{sec:solutions}

We now extend the original picket fence treatment to irradiated
atmospheres like that of close-in extrasolar giant planets. This is
achieved by linking the shortwave transfer solution to the longwave
picket fence model solution via the radiative equilibrium condition.

We first consider the radiative transfer equation as it applies to the
shortwave frequencies. The frequency integrated transfer equation for the
shortwave band, where the source function contains only a non-negligible
contribution due to coherent isotropic scattering, is given by
\begin{equation}
  \mu\der{I_s}{\tau} = \frac{\gamma}{\ves}[I_s - (1 - \ves)J_s],
\end{equation}
with
\begin{equation}
  \gamma = \frac{\kappa_s}{\kappa_c}, \quad
  \ves = \frac{\kappa_s}{k_s},
\end{equation}
where $\kappa_s$ and $k_s$ are the shortwave absorption and extinction
opacities respectively.
The parameter $\gamma$ quantifies the strength of the shortwave opacity to
its longwave continuum counterpart and $\ves$ measures the fraction of
shortwave photons lost to absorption.
The moment equations for the shortwave are
\begin{equation}
\label{eq:firstmom-irr}
  \der{H_s}{\tau} = \gamma J_s,
\end{equation}
and
\begin{equation}
\label{eq:secmom-irr}
  \der{K_s}{\tau} = \frac{\gamma}{\ves} H_s,
\end{equation}
Using the closure relation $K_s = \bar{\mu}^2 J_s$
\citep{guillot2010,hengetal2012,hengetalrt2014}
where $\bar{\mu}$ is the cosine of the angle of the collimated stellar beam
with respect to the vertical, we obtain the second order ordinary differential
equation
\begin{equation}
\label{eq:2ode-Js}
  \dder{J_s}{\tau} = \frac{\gamu^2}{\ves}J_s,
\end{equation}
which has the simple exponential solution
\begin{equation}
\label{eq:Js-sol}
  J_s = J_s(0)\exp\left( -\gamus\tau \right),
\end{equation}
where $\gamu = \gamma/|\bar{\mu}|$ and $\gamus = \gamu/\sqrt{\ves}$ are
assumed constant.
Consistency with Equation~(\ref{eq:firstmom-irr}) implies
\begin{equation}
  H_s = H_s(0)\exp\left( -\gamus\tau \right)
\end{equation}
so that $H_s = \bar{\mu}\sqrt{\ves}J_s$ \citep{hengetalrt2014}.

The radiative equilibrium condition which is given by $d(H_1 + H_2 + H_s)/d\tau = 0$
implies
\begin{equation}
\label{eq:radeq12-irr}
  \lambda J_1 + J_2 + \gamma J_s = [\lambda \beta + 1 - \beta]\tB,
\end{equation}
Adding Equations~(\ref{eq:firstmom-1}) and (\ref{eq:firstmom-2}), we have
\begin{equation}
  \der{}{\tau}(H_1 + H_2) = -\gamma J_s = \gamus H_s,
\end{equation}
which has the full solution
\begin{equation}
\label{eq:Fl}
  H_1 + H_2 = \calh - H_s.
\end{equation}
where $\calh$ is the total integrated longwave Eddington flux.
Combining Equations~(\ref{eq:secmom-1}) and (\ref{eq:secmom-2}) we obtain
\begin{equation}
  \der{}{\tau}\left( \frac{K_1}{\eta} + K_2 \right) =
  \frac{\calh}{\vecm} - \frac{H_s}{\vecm},
\end{equation}
which has the full solution
\begin{equation}
\label{eq:Kl}
  \frac{K_1}{\eta} + K_2 = \frac{\calh}{\vecm}\tau + \frac{c}{\vecm}
  + \frac{H_s}{\vecm\gamus}.
\end{equation}
With the Eddington approximation $J_{1, 2} = 3K_{1, 2}$, we may express
the integrated Planck function as
\begin{equation}
  \tB = \frac{3}{\tl}
  \left[ \lambda K_1 + K_2 - \frac{\gamus}{3}H_s \right].
\end{equation}
Using Equation~(\ref{eq:Kl}) this may be written in either of the two forms
given below
\begin{eqnarray}
\label{eq:curlyB-K1}
  \tB &=& \frac{3}{\tl}
  \left[ \frac{\calh}{\vecm}\tau + \frac{c}{\vecm} + \left( \lambda - \frac{1}
  {\eta}
  \right)K_1
  + \left( \frac{1}{\vecm\gamus} - \frac{\gamus}{3} \right)H_s
   \right],  \\
\label{eq:curlyB-K2}
  \tB &=& \frac{3}{\tl}
  \left[ \lambda\eta\left(\frac{\calh}{\vecm}\tau + \frac{c}{\vecm} \right) + (1 -
  \lambda\eta)K_2
  + \left( \frac{\lambda\eta}{\vecm\gamus} - \frac{\gamus}{3} \right)H_s
  \right]. \nonumber \\
\end{eqnarray}
Combining Equation~(\ref{eq:secmom-1}) with Equation~(\ref{eq:firstmom-1})
and Equation~(\ref{eq:secmom-2}) with Equation~(\ref{eq:firstmom-2})
by using the Eddington approximation, we obtain the pair of
inhomogeneous second order ordinary differential equations
\begin{eqnarray}
\label{eq:2ode-K1-gen}
  \dder{K_1}{\tau} &=& \frac{\eta\lambda}{\vecm}(3K_1 - \beta\tB), \\
\label{eq:2ode-K2-gen}
  \vecm\dder{K_2}{\tau} &=& 3K_2 - (1 - \beta)\tB.
\end{eqnarray}
Substituting Equations~(\ref{eq:curlyB-K1}) and (\ref{eq:curlyB-K2}) in
Equations~(\ref{eq:2ode-K1-gen}) and (\ref{eq:2ode-K2-gen}) yields
\begin{eqnarray}
\label{eq:2ode-K1-irr}
  \dder{K_1}{\tau} &=&
  \frac{3\lambda\teta}{\vecm\tl}K_1
  - \frac{3\eta\lambda\beta}{\vecm^2\tl}(\calh\tau + c)
  \nonumber \\
  &\phantom{==}&
  - \frac{3\eta\lambda\beta}{\vecm\tl}
  \left( \frac{1}{\vecm\gamus} - \frac{\gamus}{3} \right)H_s,  \\
\label{eq:2ode-K2-irr}
  \dder{K_2}{\tau} &=&
  \frac{3\lambda\teta}{\vecm\tl}K_2
  - \frac{3\eta\lambda(1 - \beta)}{\vecm^2\tl}(\calh\tau+c)
  \nonumber \\
  &\phantom{==}&
  - \frac{3(1 - \beta)}{\vecm\tl}
  \left( \frac{\lambda\eta}{\vecm\gamus} - \frac{\gamus}{3} \right)H_s,
\end{eqnarray}
where we have defined the convenient shorthands
\begin{eqnarray}
  \tl &=& \lambda \beta + 1 - \beta, \\
  \teta &=& \beta + \eta(1 - \beta).
\end{eqnarray}
Bounded solutions to Equations~(\ref{eq:2ode-K1-irr}) and
(\ref{eq:2ode-K2-irr}) are given by
\begin{eqnarray}
\label{eq:K1-sol-irr}
  K_1 &=& a \exp(-q\tau)
  + \frac{\eta\beta}{\vecm\teta}(\calh\tau + c)
  \nonumber \\  &\phantom{==}&
  - \frac{3\lambda\eta\beta}{\gamus^2\vecm\tl - 3\lambda\teta}
  \left( \frac{1}{\vecm\gamus} - \frac{\gamus}{3} \right)H_s,  \\
\label{eq:K2-sol-irr}
  K_2 &=& b \exp(-q\tau)
  + \frac{\eta(1 - \beta)}{\vecm\teta}(\calh\tau + c)
  \nonumber \\ &\phantom{==}&
  - \frac{3(1 - \beta)}{\gamus^2\vecm\tl - 3\lambda\teta}
  \left( \frac{\lambda\eta}{\vecm\gamu} - \frac{\gamus}{3} \right)H_s,
\end{eqnarray}
where $q$, its inverse rather, is a characteristic optical depth that is
given by
\begin{equation}
\label{eq:q}
  q = \sqrt{\frac{3\lambda\teta}{\vecm\tl}}.
\end{equation}
In order that Equations~(\ref{eq:K1-sol-irr}) and (\ref{eq:K2-sol-irr})
add up to Equation~(\ref{eq:Kl}), we require $a = -b\eta$.
Finally, the integrated Planck function may be expressed as
\begin{eqnarray}
  \tB &=& \frac{3\eta}{\vecm\teta}(\calh\tau + c)
  -\frac{3(\lambda\eta - 1)}{\tl}b\exp(-q\tau)
  \nonumber \\ &\phantom{==}&
  - \frac{(\gamus^2\vecm - 3)(\gamus^2\vecm - 3\eta\lambda)}
  {\gamus\vecm(\gamus^2\vecm\tl - 3\lambda\teta)}H_s(0)\exp(-\gamus\tau),
  \nonumber \\
\end{eqnarray}

Thus it remains only to determine the integration constants $b$ and $c$ which
we achieve by the application of suitable boundary conditions.
For the sake of conformity with \citet{chandra35}, we use the relations
$\calf = 4\calh$ and $\Fs = 4H_s(0)$.
Assuming that at $\tau = 0$, we have $4 H_{1,2}(0) = 6 K_{1,2}(0)$
\citep{chandra35} and by considering Equations~(\ref{eq:secmom-1}) and
(\ref{eq:secmom-2}) evaluated at $\tau = 0$, we obtain
\begin{eqnarray}
\label{eq:bc-1}
  (6\eta + 4q)b - \frac{6\eta\beta}{\vecm\teta}c + \frac{\beta}
  {\vecm\teta}\calf &&
  \nonumber \\
  + \frac{3\lambda\beta\left(\gamus + \frac{3}{2}\eta\right)}
  {\gamus^2\vecm\tl - 3\lambda\teta}
  \left( \frac{1}{\vecm\gamus} - \frac{\gamus}{3} \right)\Fs &=& 0,  \\
\label{eq:bc-2}
  (6 + 4q)b + \frac{6\eta(1 - \beta)}{\vecm\teta}c -
  \frac{\eta(1 - \beta)}{\vecm\teta}\calf  &&
  \nonumber \\
  - \frac{3(1 - \beta)\left(\gamus + \frac{3}{2}\right)}
  {\gamus^2\vecm\tl - 3\lambda\teta}
  \left( \frac{\lambda\eta}{\vecm\gamus} - \frac{\gamus}{3} \right)\Fs
  &=& 0.
\end{eqnarray}
The boundary condition used here corresponds to the application of
the second and third possibility listed in \citet[see their Section 2.3.2]
{pg2014} with the second Eddington coefficient $f_{\mathrm{H}} = 1/2$.
Solving the system of equations~(\ref{eq:bc-1}) and (\ref{eq:bc-2}) results in
\begin{equation}
  b = b_l \calf + b_s \Fs, \quad
  c = c_l \calf + c_s \Fs,
\end{equation}
where
\begin{eqnarray}
  b_l &=& \frac{\beta(1 - \beta)(\eta - 1)}{\vecm\teta(6\teta + 4q)},  \\
  b_s &=&
  \frac{\beta(1 - \beta)
  [\gamus^2\vecm(\lambda - 1) + 3\lambda(\eta - 1)]}
  {\vecm(\gamus^2\vecm\tl - 3\lambda\teta)(6\teta + 4q)}
  \nonumber \\ &\phantom{==}&
  + \frac{3}{2}\frac{\beta(1 - \beta)\gamus(\lambda\eta - 1)}
  {(\gamus^2\vecm\tl - 3\lambda\teta)(6\teta + 4q)},
  \\
  c_l &=& \frac{6[\beta + \eta^2(1 - \beta)] + 4q\teta}{6\eta(6\teta + 4q)},
  \\
  c_s &=&
  \frac{\teta
  \left( \gamus + \frac{3}{2} \right)
  (6\eta + 4q)[\gamus^2\vecm(\beta - 1) + 3\lambda\eta]}
  {6\gamus\eta(\gamus^2\vecm\tl - 3\lambda\teta)(6\teta + 4q)}
  \nonumber \\ &\phantom{==}&
  - \frac{3\teta\lambda\beta(\eta - 1)
  [9\eta + \gamus(6 + 4q + 6\eta)]
  }{6\gamus\eta(\gamus^2\vecm\tl - 3\lambda\teta)(6\teta + 4q)}
  \nonumber \\ &\phantom{==}&
    - \frac{\teta\lambda\beta\gamus\vecm(6 + 4q)
  \left( \gamus + \frac{3}{2}\eta \right)
  }{6\eta(\gamus^2\vecm\tl - 3\lambda\teta)(6\teta + 4q)}.
\end{eqnarray}

We now express the fluxes in terms of their respective equilibrium
temperatures as given by
\begin{equation}
  \tB = \frac{\sigma_{\mathrm{SB}} T^4}{\pi}, \quad
  \calf = \frac{\sigma_{\mathrm{SB}} \Tint^4}{\pi}, \quad
  \Fs = \bar{\mu}\frac{\sigma_{\mathrm{SB}} \Tirr^4}{\pi},
\end{equation}
where $\sigma_{\mathrm{SB}}$ is the Stefan-Boltzmann constant, $\Tint$ is the
internal temperature associated with the thermal flux at the bottom of the
atmospheres and $\Tirr$ is the irradiation temperature associated with the flux
at the top of the atmosphere.
This substitution results in the temperature profile given by
\begin{eqnarray}
\label{eq:tprofile}
  T^4 &=& \frac{3\Tint^4}{4}\left[ \frac{\eta}{\vecm\teta}\tau
  + \frac{4\eta}{\vecm\teta}c_l
  - \frac{4(\eta\lambda - 1)}{\tl}b_l\exp(-q\tau) \right]
  \nonumber \\
  &\phantom{=}& - \frac{3|\bar{\mu}|\Tirr^4}{4}
  \left[ \frac{4\eta}{\vecm\teta}c_s
  - \frac{4(\eta\lambda - 1)}{\tl}b_s\exp(-q\tau)
  \right.
  \nonumber \\ &\phantom{==}&
  \left.
  - \frac{(\gamus^2\vecm - 3)(\gamus^2\vecm - 3\eta\lambda)}
  {3\gamus\vecm(\gamus^2\vecm\tl - 3\lambda\teta)}\exp(-\gamus\tau)
  \right].
\end{eqnarray}
In the limit of $\beta \to 0$ and no external irradiation $\Tirr = 0$,
we have
\begin{equation}
  \teta = \eta, \quad \tl = 1, \quad b_l = 0, \quad c_l = \frac{1}{6},
\end{equation}
and equation~(\ref{eq:tprofile}) reduces to the classic Milne's solution
\begin{equation}
  T^4 = \frac{3}{4}\Tint^4\left(\tau + \frac{2}{3} \right).
\end{equation}
With only $\Tirr = 0$, we recover the non-irradiated solution of
\citet[see Equation (52)]{chandra35}.

\section{Results}
\label{sec:discussion}

The picket fence model of \citet{chandra35} was originally developed to
capture the effects that arise due to line blanketing in stellar atmosphere
models.
The initial treatment was based on the ideal case of lines with uniform
width, strength and separation.
A discussion of the specific limitations that result from these assumptions
is presented in \citet{as1969,athaybook}.
Nevertheless, the original model did succeed in illustrating the basic
effects due to line blanketing, namely that of cooler surface temperatures
and warmer deeper temperatures. We have, therefore, retained the same
simplified treatment of the lines in order to
construct a general non-gray model that includes coherent isotropic
scattering. All the results presented here are derived assuming fiducial
values for the internal temperature $\Tint = 100 \, K$, the irradiation
temperature $\Tirr = 1000 \, K$ and irradiation angle given by
$|\bar{\mu}| = 1/\sqrt{3}$.

\subsection{Surface Cooling and Backwarming}

\begin{figure}
  \centering
  \includegraphics{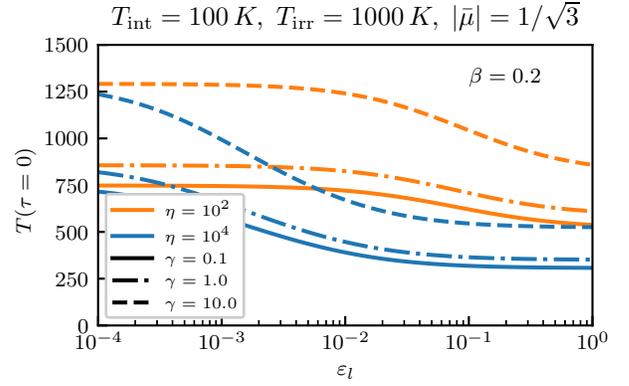}
  \caption{Boundary temperature $T(\tau = 0)$ of an irradiated
  atmosphere as a function of the line scattering parameter $\vel$
  at a fixed line width $\beta$ but different values of the line
  strength $\eta$ and shortwave absorption opacity parameter $\gamma$.}
  \label{fig:skintemp}
\end{figure}

\begin{figure*}
  \centering
  \includegraphics{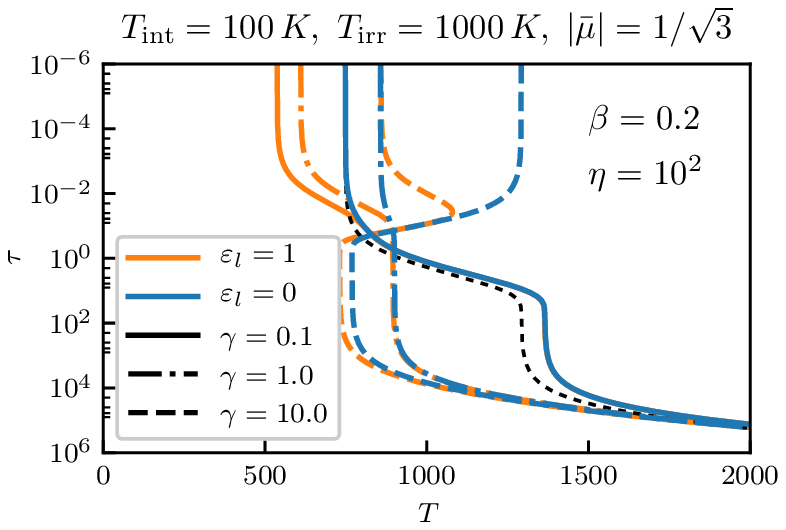}
  \includegraphics{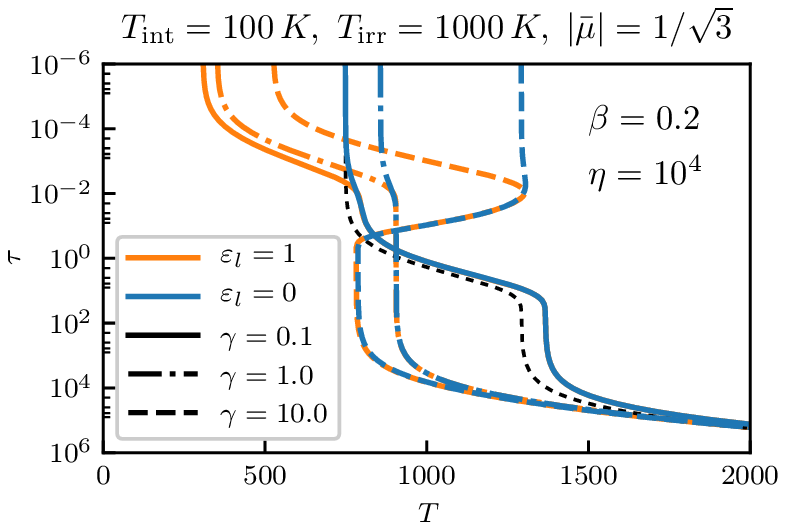}
  \caption{Temperature profiles
  for a given line width $\beta$ and two different values of the line
  strength $\eta$.
  The temperature profile in each case is derived for three different values
  of the shortwave opacity parameter $\gamma$ represented by the solid,
  dashed and dash-dotted curves.
  The blue curves represent the profile with lines that are due to pure
  scattering whereas the orange curves represent lines due to pure absorption.
  $\gamma > 1$ results in an anti-greenhouse effect leading to a
  relatively hotter upper atmosphere and cooler lower atmosphere.
  The dotted black line is the gray temperature profile computed for
  $\gamma = 0.1$ and is plotted here to illustrate the backwarming effect
  seen in the solid (blue and orange) curves for the same $\gamma$.}
  \label{fig:tp2}
\end{figure*}

Line blanketing introduces two main effects in the thermal structure and
spectrum of atmospheres\footnote{Some works include the additional effect of line \emph{blocking} which we do not consider here. Our usage of the term blanketing is in the sense of \citet{athaybook}.}.
If line formation can be attributed solely to absorption processes, the
temperature of the upper layer is lower compared to what it would be
in the absence of lines (the gray limit), an effect that is referred to as \emph{surface cooling} \citep{athaybook}.
The degree of surface cooling is lowered if the lines are partly
due to scattering processes parameterized here by $\vel$.
The surface cooling is completely absent if the lines are entirely due to
scattering $\vel \to 0$.
Figure~\ref{fig:skintemp} shows the change in boundary or skin temperature,
here referred to as the temperature at zero optical depth,
with the longwave scattering parameter $\vel$ for a fixed line width but
different line strengths.
In the case of irradiated atmospheres, the strength of the shortwave opacity also influences the boundary temperature due to greater or lower relative absorption of incident starlight in the upper layers. Higher values of $\gamma$ therefore lead to higher upper layer temperatures.

\begin{figure}
  \centering
  \includegraphics{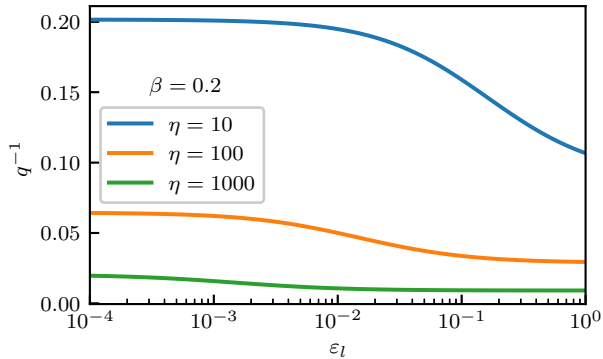}
  \caption{Characteristic optical depth $q^{-1}$, below which surface
  cooling effects are observed, as a function of the line scattering parameter
  $\ve_l$ for different values of the line strength $\eta$ at a fixed line
  width $\beta$.}
  \label{fig:qinve}
\end{figure}

The deeper layer warming observed in the non-gray model that
we have derived here is due to the effect of line blanketing. This
\emph{backwarming} results from the increase in temperature due to an
attendant increase in the radiative flux per unit interval that is redistributed within the continuum band of frequencies as a result of its occlusion by the lines. Deeper layer backwarming is, however, insensitive to the nature of line formation.
Figure~\ref{fig:tp2} shows temperature profiles for two limiting values of the scattering parameter $\vel$ and two different values of the line strength $\eta$.
The profiles are calculated for a fixed angle of the irradiation beam, a
constant line width, fiducial values of the effective internal and irradiation temperatures and three different values of the shortwave absorption opacity (excluding the effect of shortwave scattering here).
We plot the gray temperature profile with $\gamma = 0.1$ as a dotted black line in Figure~\ref{fig:tp2} to illustrate the backwarming
effect. Notice that the corresponding non-gray temperature profile with $\vel = 1$ and $\vel = 0$ in Figure~\ref{fig:tp2} are both warmer by the same extent with reference to the dotted line thereby illustrating the the deeper layer warming as well as its insensitivity to the nature of line formation.
The degree of backwarming increases with the width of the line, represented here by $\beta$ and has been examined in detail by \citet{pg2014}.
In the limit $\vel = 1$ and $\gamma \gtrsim 1$, we see, as also found in \citet{pg2014}, that the lower boundary temperatures confine the heating due to stronger shortwave absorption into a thin hot layer immediately below the surface. This thin hot layer is of course absent in the opposite limit of $\ve_l = 0$ due to the lack of surface cooling.

\subsection{Limit optical depth}

\begin{figure*}
  \centering
  \includegraphics{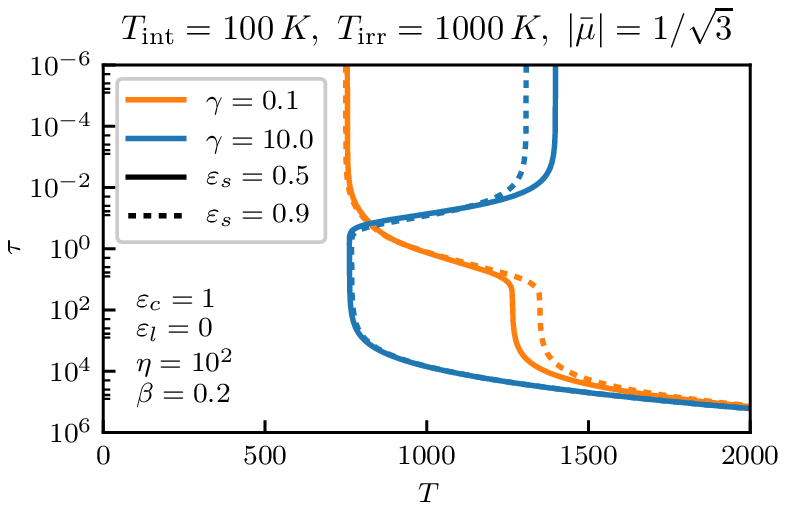}
  \includegraphics{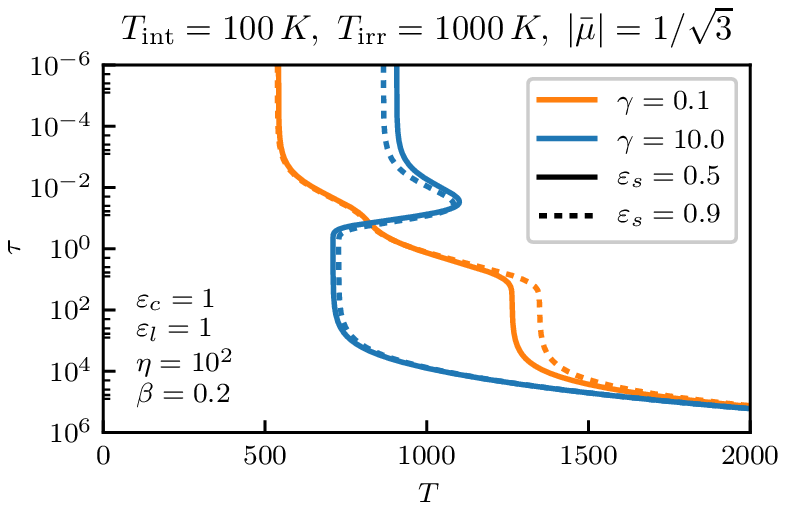}
  \caption{Temperature profiles
  for two different values of the shortwave continuum coherent isotropic
  scattering parameter $\ves$, represented by the solid and dotted lines. The
  profiles are derived in the presence of spectral lines of a
  given width and strength as well as two different values of the shortwave
  opacity parameter $\gamma$ represented by the orange and blue curves.
  The left panel illustrates the temperature profile
  when the lines are due to pure scattering and the right panel is for lines
  due to pure absorption.
  We consider the limit of pure absorption in the longwave continuum here.
  The inclusion of coherent shortwave continuum scattering results in a
  leftward shift in the temperature profile towards cooler temperatures.}
  \label{fig:sws2}
\end{figure*}

\begin{figure*}
  \centering
  \includegraphics{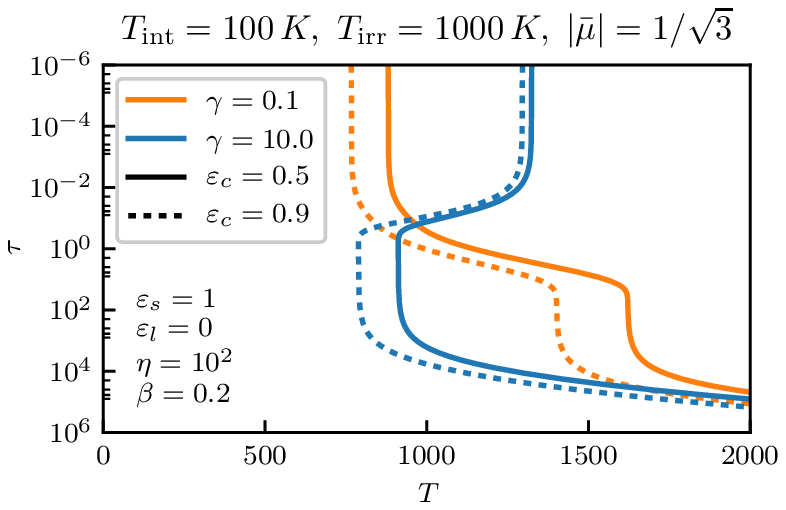}
  \includegraphics{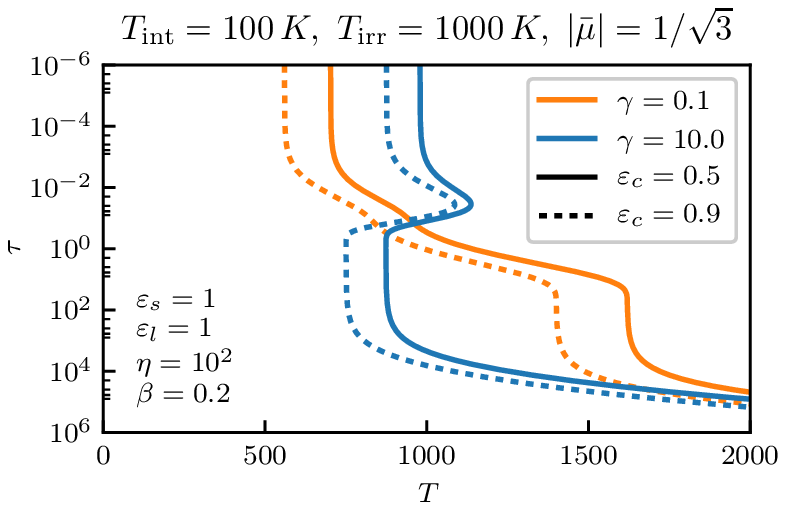}
  \caption{Temperature profiles
  for two different values of the longwave continuum coherent isotropic
  scattering parameter $\vecm$, represented by the solid and dotted lines.
  The profiles are derived in the presence of spectral lines of a
  given width and strength as well as two different values of the shortwave
  opacity parameter $\gamma$ represented by the orange and blue curves.
  The left panel illustrates the temperature profile
  when the lines are due to pure scattering and the right panel displays the
  profiles when the lines are due to pure absorption.
  We consider the limit of pure absorption in the shortwave continuum here.
  The inclusion of coherent longwave continuum scattering results in a
  rightward shift in the temperature profile towards hotter temperatures.}
  \label{fig:lws2}
\end{figure*}
\textbf{}
A characteristic depth that emerges from the picket fence calculation
is given in terms of $q^{-1}$ as defined by equation~(\ref{eq:q}).
The corresponding quantity in \citet{pg2014} is referred to as $\tau_{\mathrm{lim}}$ and it represents the depth above which the surface cooling
effect may be present in the atmosphere provided that the lines are due in
some part to absorption processes.
This scale is a function of the line width $\beta$, the line opacity ratio $\xi$ as well as the scattering parameters $\vel$ and $\vecm$.
However, as shown in Figure~\ref{fig:qinve}, the variation of $q^{-1}$ with respect to $\vel$ is negligible and is largely decided by the line width and strength.
Similarly, any meaningful change in $\vecm$ brings about only a
negligible modification to $q^{-1}$.
Our results derived for $\vel \to 1$ are qualitatively similar to
\citet{pg2014} and differ quantitatively only by a factor of a Rosseland mean
opacity $\varrho$, defined here in dimensionless form as
\citep[see Equation (56)]{chandra35}
\begin{equation}
  \frac{1}{\varrho} \equiv \frac{\beta}{\eta} + 1- \beta.
\end{equation}

\subsection{Longwave and Shortwave continuum scattering}

The effect of coherent scattering in the shortwave is to push the temperature
profile to lower values on a near global scale \citep{hengbook}.
Coherent isotropic continuum scattering in the shortwave is parameterized in terms of $\ves$ and its effect on the temperature profile is demonstrated in Figure~\ref{fig:sws2}.
The global shift in the thermal profile towards lower temperatures adds to any
surface cooling present that is due to absorption lines and also effectively
offsets the backwarming in the mid to deeper layers. The lower temperatures
result from a reduction in the total energy budget by a factor of $1 - A_{
\mathrm{B}}$ where $A_{\mathrm{B}}$ is the Bond albedo which may be expressed
in terms of the scattering parameter $\ves$ \citep{hengetal2012}.

Coherent isotropic scattering in the longwave continuum band of frequencies has a similar atmosphere-wide effect where the temperatures are now shifted to higher values as illustrated in Figure~\ref{fig:lws2}.
This is a manifestation of the classical greenhouse effect
\citep{hengbook} and is different from the lack of surface cooling due to
lines formed by scattering.
The former is an actual warming process and is present on a global scale whereas the latter is the result of the lines being uncoupled from the thermal energy reservoir \citep{ml1971}. Taken together, scattering processes therefore play an important role in determining the equilibrium temperature profile even in simple pseudo-non-gray models.

\section{Summary}
\label{sec:summary}

We have derived an analytical model for irradiated atmospheres that
combines the effect of spectral lines in the longwave band of frequencies,
where the lines may be due to either pure absorption or pure coherent
scattering processes or some combination of the two.
To achieve this, we adapted the picket-fence treatment of \citet{chandra35}
to model line blanketing effects. The picket-fence treatment has been recently used to model irradiated atmospheres but without including the possibility of lines due to coherent scattering \citep{pg2014}. Our results demonstrate that the cooling of the upper layers due to line blanketing depends on the nature of line formation as was previously observed in the context of non-irradiated atmospheres. If scattering is solely responsible for the lines,
then the surface temperatures retain their gray value as the lines are
not coupled to the thermal energy of the gas in this limit.
Transit spectroscopy of exoplanets is generally most sensitive to very low
pressure levels or equivalently the upper layers of the exoplanet's atmosphere \citep{madhu2014}.
Given that the surface temperature is sensitive to the line formation
process as revealed by the picket fence analysis, one must exercise caution in the interpretation of observations on the basis of atmospheric transfer models.
Furthermore, the contribution of coherent scattering in the continuum can
significantly alter global temperature levels depending on the wavelength
band.
If the planet reflects some fraction of its incident light, the deeper layer temperatures are lowered and negates the backwarming effect due to the lines.
If coherent scattering is present in the longwave continuum, the
greenhouse effect comes into play leading to greater warming throughout the atmosphere.
Our analytical model therefore accommodates a greater range of possibilities over a larger parameter space and may be used to derive reasonable estimates of the thermal structure of irradiated atmospheres.

\acknowledgments
GM thanks the Center for Space and Habitability in Bern,
where part of this work was carried out, for their hospitality.
GM thanks the Kavli Institute for Theoretical Physics,
where this work was completed, for their hospitality and support
through the KITP graduate fellowship program.
The research leading to these results has received
funding from the European Research Council under the
European Unions Seventh Framework Programme (FP/2007-
2013) under ERC grant agreement 306614.
KH thanks the Swiss National Science Foundation, the Center
for Space and Habitability, the European Research Council,
and the Swiss-based MERAC Foundation for partial financial support.
This research was supported in part by the National Science Foundation
under Grant No. NSF PHY 17-48958.

\bibliographystyle{aasjournal}

\end{document}